# DEGREES OF EQUIVALENCE IN A KEY COMPARISON

[1]*Thang H. L., Nguyen D. D.*
*Vietnam Metrology Institute,*
*Address: 8 Hoang Quoc Viet, Hanoi, Vietnam*

**Abstract:**

In an interlaboratory key comparison, a data analysis procedure for this comparison was proposed and recommended by CIPM [1, 2, 3], therein the degrees of equivalence of measurement standards of the laboratories participated in the comparison and the ones between each two laboratories were introduced but a corresponding clear and plausible measurement model was not given. Authors in [4] offered possible measurement models for a given comparison and a suitable model was selected out after rigorous analyzing steps for expectation values of these degrees of equivalence. The systematic laboratory-effects model was then selected as a right one in this report. Those models were all based on the one true value existence assumption. However in the year 2008, a new version of the Vocabulary for International Metrology (VIM) [7] was issued where the true value of a given measurement standard should be now perceived as multiple true values which following a given statistics distribution. Applying this perception of true values of a measurement standard with combination of the steps in [4], measurement models have been developed and degrees of equivalence have been analyzed. The results show that although with new definition, the systematic laboratory-effects model is still the reasonable one in a given key comparison.

## I. Introduction

In reference [2], concept of degrees of equivalence between laboratories was stated as one of important criteria in Mutual Recognition Arrangement (MRA) between National Metrology Institutes (NMIs). Degrees of equivalence are defined in [1] as following:

**Degree of equivalence of a measurement standard**: the degree to which the value of a measurement standard is consistent with the key comparison reference value. This is expressed quantitatively by the deviation from the key comparison reference value and the uncertainty of this deviation. The degree of equivalence between two measurement standards is expressed as the difference between their respective deviations from the key comparison reference value and the uncertainty of this difference.

Mathematically, the degree is expressed as $d_i = x_i - x_K$ and $u^2(d_i) = u^2(x_i) - u^2(x_K)$. The degree of equivalence between two measurement standards is expressed as $d_{ij} = x_i - x_j$ and $u^2(d_{ij}) = u^2(x_i) + u^2(x_j)$ [3]. To illuminate the statistics natures of those quantities, measurement models for a key comparison have been offered and analyzed in [4]. In those models, a given measurement standard is assumed having only one true value. Actually, as discussed in [7], a more general view should be of understanding that for a given measurement standard, there exist a set of true values which we then assume following a given statistics distribution.

---
[1] Email: thanglh@vmi.gov.vn

## II. Mathematical modeling

Let consider a given key comparison where a measurement quantity having a set of true values $Y_i$, $i = 1$ to $N$ ($N$ is the number of participants) which is following a unique stable distribution during the comparison time. The expectation and variance of $Y_i$ will be $E(Y_i) = Y$ and $V(Y_i) = s^2(Y_i)$. Call $X_1$, $X_2 \ldots X_N$ and $x_1$, $x_2 \ldots x_N$ are expectation values and measured values of the measurement quantity measured and provided by the $i^{th}$ laboratory.

Each measured value will have a reliable measurement uncertainty $u(x_i)$. Call $b_1 = (X_1 - Y_1)$, $b_2 = (X_2 - Y_2), \ldots, b_N = (X_N - Y_N)$. The set of $b_1$, $b_2, \ldots, b_N$ are not always zero due to some unrecognizable errors during the measurement but all of the measurement values of a certain laboratory should still have the same expectation value. Next some measurement models with different assumptions will be developed and their analysis will be carried out.

1. None laboratory effect

In this case the measurement equation will be of the form:

$$x_i = Y_i + e_i \qquad (1)$$

The equation for expectation values will be: $E(x_i) = X_i = Y$. Here $b_i = 0$ implies the participating laboratory makes no errors on the measurement or all the errors were recognizable and corrected. The corresponding variance equation will be:

$$V(x_i) = V(Y_i) + V(e_i) \text{ or } V(x_i) = s^2(Y_i) + u^2(e_i) \qquad (2)$$

2. Random laboratory effect

The measurement equation will be:

$$x_i = Y_i + b_i + e_i \qquad (3)$$

The expectation equation:

$$E(x_i) = E(Y_i) + E(b_i) + E(e_i) \text{ or } E(x_i) = Y \qquad (4)$$

where $b_i$ is assumed to follow a statistics distribution with zero expectation.

The variance equation:

$$V(x_i) = V(Y_i) + V(b_i) + V(e_i) \text{ or } V(x_i)$$
$$= s^2(Y_i) + s^2(b_i) + u^2(e_i) \qquad (5)$$

3. Systematic laboratory effect

The measurement equation will be:

$$x_i = Y_i + b_i + e_i \qquad (6)$$

where $b_i$ becomes a constant now.

The expectation and variance equation:

$$E(x_i) = E(Y_i) + E(b_i) + E(e_i) \quad (7)$$

$$V(x_i) = V(Y_i) + V(b_i) + V(e_i) \quad (8)$$

or

$$E(x_i) = Y + b_i \quad (9)$$

$$V(x_i) = V(Y_i) + u^2(e_i) \quad (10)$$

## III. Key reference values

### 1. None laboratory effect

The key reference value:

$$x_K = (\Sigma_i x_i / (s^2(Y_i) + u^2(e_i)))/(\Sigma_i 1/(s^2(Y_i) + u^2(e_i))),$$

$$u(x_K) = 1/\sqrt{(\Sigma_i 1/(s^2(Y_i) + u^2(e_i)))} \quad (11)$$

### 2. Random laboratory effect

The key reference value:

$$x_K = (\Sigma_i x_i / (s^2(Y_i) + s^2(b_i) + u^2(e_i)))/(\Sigma_i 1/(s^2(Y_i) + s^2(b_i) + u^2(e_i))),$$

$$u(x_K) = 1/\sqrt{(\Sigma_i 1/(s^2(Y_i) + s^2(b_i) + u^2(e_i)))} \quad (12)$$

### 3. Systematic laboratory effect

The key reference value:

$$x_K = (\Sigma_i x_i / (s^2(Y_i) + u^2(e_i)))/(\Sigma_i 1/(s^2(Y_i) + u^2(e_i))),$$

$$u(x_K) = 1/\sqrt{(\Sigma_i 1/(s^2(Y_i) + u^2(e_i)))} \quad (13)$$

## IV. Degrees of equivalence

### 1. None laboratory effect

Measurement models of any two participating laboratories:

$$x_i = Y_i + e_i \text{ and } x_j = Y_j + e_j \quad (14)$$

Deviation of measured values of two laboratories:

$$d_{ij} = x_i - x_j = Y_i - Y_j + e_i - e_j \quad (15)$$

Deviation of a measured value and the key reference value:

$$d_i = x_i - x_K = Y_i + e_i - (\Sigma_j x_j / (s^2(Y_j) + u^2(e_j)))/(\Sigma_j 1/(s^2(Y_j) + u^2(e_j))) \quad (16)$$

The expectation values:

$$E(d_{ij}) = E(Y_i) - E(Y_j) + E(e_i) - E(e_j)$$

$$= 0,$$

$$E(d_i) = E(x_i) - E(x_K)$$

$$= E(Y_i) + E(e_i) - (\Sigma_j E(x_j)/ (s^2(Y_j) + u^2(e_j)))/(\Sigma_j 1/(s^2(Y_j) + u^2(e_j)))$$

$$= E(Y_i) + E(e_i) - (\Sigma_j E(Y_j + e_j)/ (s^2(Y_j) + u^2(e_j)))/(\Sigma_j 1/(s^2(Y_j) + u^2(e_j)))$$

$$= E(Y_i) + E(e_i) - (\Sigma_j E(Y_j) / (s^2(Y_j) + u^2(e_j))) / (\Sigma_j 1/ (s^2(Y_j) + u^2(e_j)))$$

$$= E(Y_i) - (\Sigma_j E(Y_j) / (s^2(Y_j) + u^2(e_j))) / (\Sigma_j 1/ (s^2(Y_j) + u^2(e_j)))$$

$$= E(Y_i) - E(Y_j)$$

$$= 0 \tag{17}$$

## 2. Random laboratory effect

Measurement models of any two participating laboratories:

$$x_i = Y_i + b_i + e_i \text{ and } x_j = Y_j + b_j + e_j \tag{18}$$

Deviation of measured values of two laboratories:

$$d_{ij} = x_i - x_j = Y_i - Y_j + b_i - b_j + e_i - e_j \tag{19}$$

Deviation of a measured value and the key reference value:

$$d_i = x_i - x_K$$

$$= Y_i + b_i + e_i - (\Sigma_j x_j / (s^2(Y_j) + s^2(b_j) + u^2(e_j))) / (\Sigma_j 1/ (s^2(Y_j) + s^2(b_j) + u^2(e_j))) \tag{20}$$

The expectation values:

$$E(d_i) = E(Y_i) + E(b_i) + E(e_i) - (\Sigma_j E(x_j)/ (s^2(Y_j) + s^2(b_j) + u^2(e_j))) / (\Sigma_j 1/ (s^2(Y_j) + s^2(b_j) + u^2(e_j)))$$

$$= E(Y_i) + E(b_i) + E(e_i) - E(Y_j) = 0 \text{ and } d_{ij} = 0 \tag{21}$$

## 3. Systematic laboratory effect

Measurement models of any two participating laboratories:

$$x_i = Y_i + b_i + e_i \text{ và } x_j = Y_j + b_j + e_j \tag{22}$$

Deviation of measured values of two laboratories:

$$d_{ij} = x_i - x_j = Y_i - Y_j + b_i - b_j + e_i - e_j \tag{23}$$

Deviation of a measured value and the key reference value:

$$d_i = x_i - x_K$$

$$= Y_i + b_i + e_i - (\Sigma_j x_j / u^2(e_j)) / (\Sigma_j 1/ (u^2(Y_j) + u^2(e_j))) \tag{24}$$

The expectation values:

$$E(d_i) = E(Y_i) + E(b_i) + E(e_i) - (\Sigma_j E(Y_j + b_j + e_j)/ u^2(e_j)) / (\Sigma_j 1/ (u^2(Y_j) + u^2(e_j)))$$

$$= E(Y_i) + E(b_i) + E(e_i) - (\Sigma_j E(Y_j) + E(b_j) + E(e_j))/ u^2(e_j)) / (\Sigma_j 1/ (u^2(Y_j) + u^2(e_j)))$$

$$= E(Y_i) + E(b_i) + E(e_i) - E(Y_j) - \Sigma_j b_j / u^2(e_j)) / (\Sigma_j 1/ (u^2(Y_j) + u^2(e_j)))$$

$$= E(b_i) - \Sigma_j b_j / u^2(e_j)) / (\Sigma_j 1/ (u^2(Y_j) + u^2(e_j)))$$

$$= b_i - (\Sigma_j b_j / u^2(e_j)) / (\Sigma_j 1/ (u^2(Y_j) + u^2(e_j))) \text{ and }$$

$$E(d_{ij}) = E(Y_i) - E(Y_j) + E(b_i) - E(b_j) + E(e_i) - E(e_j)$$

$$= b_i - b_j \tag{25}$$

## V. Discussion

The approach in this report accepted the assumption of existence of a set of true values instead of the existence of only one unique true value for a given measurement standard of the artifact in a key comparison. Those true values are distributed in a common probabilistic density function. The corresponding degrees of equivalence, or in other words, the deviations and their measurement uncertainties are then analyzed. It is then seen that if a given participating laboratory did not contribute any error to the measurement or the error contributed of this laboratory to the measurement is random in nature as seen in equations (17) and (21), then the expectations are always zero. These imply that the laboratories under question are always equivalent which is not a reasonable acceptance. This fact implies that they should not be good models for a key comparison.

In contrast, if a participating laboratory contributed to the measurement a systematic error then the expectations of deviations are not possibly zero in all cases as seen in equation (25). The systematic errors committed by each one $b_i$ and $b_j$ and their uncertainties will definitely decide if they are equivalent or not. And then this model could be assigned to be a good model to describe the measurement process. It is worthy to notice that this conclusion is coincident to the one in [4].

## VI. Conclusion

In this report, the degree of equivalence is considered in three different models. The explicit deviations of each laboratory pairs and that of one laboratory with the key reference value are derived. The expectations of the deviations and then the degrees of equivalence are analyzed for each model with the assumption of multiple true values. The result support that the laboratory's systematic error model is the accepted one. The result is similar to the one in [4].

**Acknowledgement**: Dr. Nguyen Duc Dung and Dr. Tran Bao have contributed to the discussion and minutes of the report.